\documentclass[11pt]{article}

\usepackage[final]{acl}

\usepackage{times}
\usepackage{latexsym}

\usepackage[T1]{fontenc}
\usepackage[utf8]{inputenc}

\usepackage{microtype}

\usepackage{inconsolata}

\usepackage{graphicx}
\usepackage{amssymb}
\usepackage{tcolorbox}
\usepackage{multirow}
\usepackage{listings}
\usepackage{xcolor}
\usepackage{amsmath}   % For better mathematical support (optional, but recommended)
\usepackage{booktabs}  % For better table formatting (toprule, midrule, etc.)
\usepackage{fvextra}   % enhanced Verbatim with line breaking
\DefineVerbatimEnvironment{PromptBlock}{Verbatim}{
  fontsize=\footnotesize,
  breaklines=true,
  breakanywhere=true,
  commandchars=\\\{\},
  frame=lines,
  framesep=2mm
}

\title{Conformity Dynamics in LLM Multi-Agent Systems: The Roles of Topology and Self–Social Weighting}

\author{
  \textbf{Chen Han}\textsuperscript{1,2}, 
  \textbf{Jin Tan}\textsuperscript{3}, 
  \textbf{Bohan Yu}\textsuperscript{1,4}, 
  \textbf{Wenzhen Zheng}\textsuperscript{2,5}, 
  \textbf{Xijin Tang}\textsuperscript{1,2} \\
  \textsuperscript{1}School of Advanced Interdisciplinary Sciences, UCAS\\
  \textsuperscript{2}State Key Laboratory of Mathematical Sciences, AMSS, CAS \\
  \textsuperscript{3}School of Economics and Management, UCAS \\
  \textsuperscript{4}The Key Laboratory of Cognition and Decision Intelligence for Complex Systems, CASIA\\
  \textsuperscript{5}StepFun \\
  \texttt{\{hanchen23, tanjin23, yubohan23\}@mails.ucas.ac.cn}\\
  \texttt{zhengwenzhen@stepfun.com},  \texttt{xjtang@iss.ac.cn}
}

\begin{document}
\maketitle

\begin{abstract}
Large Language Models (LLMs) are increasingly instantiated as interacting agents in multi-agent systems (MAS), where collective decisions emerge through social interaction rather than independent reasoning. A fundamental yet underexplored mechanism in this process is conformity, the tendency of agents to align their judgments with prevailing group opinions. This paper presents a systematic study of how network topology shapes conformity dynamics in LLM-based MAS through a misinformation detection task. 
We introduce a confidence-normalized pooling rule that controls the trade-off between self-reliance and social influence, enabling comparisons between two canonical decision paradigms: Centralized Aggregation and Distributed Consensus. Experimental results demonstrate that network topology critically governs both the efficiency and robustness of collective judgments. Centralized structures enable immediate decisions but are sensitive to hub competence and exhibit same-model alignment biases. In contrast, distributed structures promote more robust consensus, while increased network connectivity speeds up convergence but also heightens the risk of wrong-but-sure cascades, in which agents converge on incorrect decisions with high confidence. These findings characterize the conformity dynamics in LLM-based MAS, clarifying how network topology and self–social weighting jointly shape the efficiency, robustness, and failure modes of collective decision-making.
The code will be publicly released upon acceptance.

% The code and dataset are available at: \href{https://anonymous.4open.science/r/Topology-of-Multi-Agent-Systems-5FF1}{\texttt{Topology-of-Multi-Agent-Systems}}.

\end{abstract}

\begin{figure*}[htbp]
\centering
\includegraphics[width=0.9\linewidth]{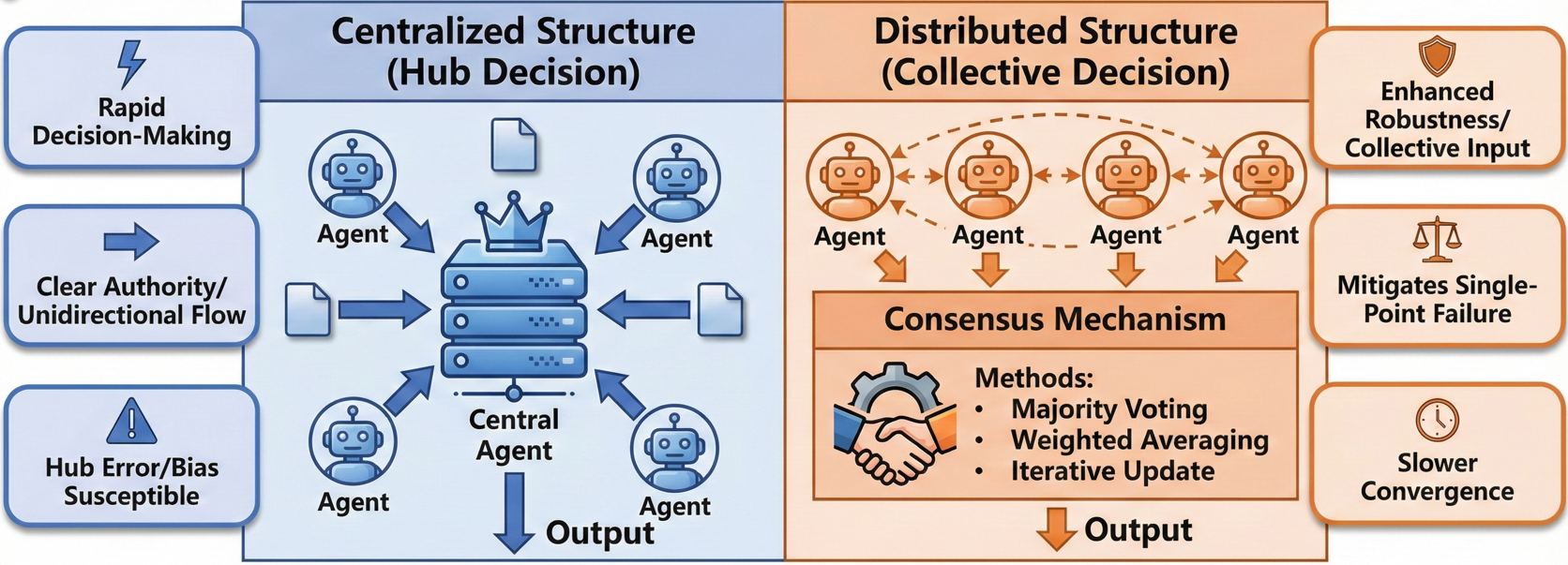}
\caption{Illustration of centralized (hub-based) and distributed (collective) multi-agent structures, highlighting their decision-making mechanisms, relative advantages, and inherent limitations in collective inference.}
\vspace{-10pt}
\label{fig:bg}
\end{figure*}

\section{Introduction}
Large Language Models (LLMs) have been increasingly instantiated as interacting agents within Multi-Agent Systems (MAS) \citep{10.5555/3709347.3743854}. Across a wide range of applications, including collaborative problem solving \citep{10.5555/3692070.3692537}, complex reasoning \citep{zhang-xiong-2025-debate4math} and misinformation detection \citep{li2025multiagentframeworkautomateddecision}, the effectiveness of collective decision-making depends not only on the competence of individual agents, but also on the social dynamics induced by their interactions \citep{ghoshal2025misinformation, cisneros-velarde-2025-biases}. A central mechanism underlying these dynamics is conformity, the tendency of agents to adjust their judgments toward majority opinions. Prior studies suggest that conformity can reduce idiosyncratic noise and facilitate coordination \citep{choi-etal-2025-empirical}. However, excessive conformity may also cause information cascades, leading groups to converge on incorrect conclusions with high confidence \citep{10.1257/jel.20241472, pinheiro_heterogeneous_2025}.

Existing research on MAS has primarily emphasized task efficacy, focusing on protocol design, role specialization, and coordination efficiency \citep{agashe-etal-2025-llm, grötschla2025}. For example, multi-agent debate frameworks can shift individual judgments toward majority positions, improving reliability while reinforcing systematic biases \citep{han2025debatetodetect}. Related work on persuasion further suggests that LLM agents tend to imitate dominant argumentative patterns during the interaction \citep{10.1073/pnas.2412815122}. While conformity has been studied in computational opinion dynamics \citep{Helfmann2023OpinionDynamics, DING2025108730, 10.1007/978-981-95-4990-0_25}, existing approaches still lack an explicit treatment of how agents generate judgments during decision processes, and in particular how interaction topology and neighbor effect jointly govern the propagation, aggregation, and amplification of agent confidence.

In this study, we bridge this gap by investigating how network topology modulates conformity through a binary misinformation detection task. We propose a confidence-normalized update rule governed by a global self-weighting parameter $\alpha$, which balances an agent's self-reliance against their neighborhood influence. Specifically, we compare two distinct decision modes: (i) \textbf{Centralized Aggregation} (e.g., star networks), where a hub synthesizes a collective decision in a single round; and (ii) \textbf{Distributed Consensus} (e.g., sparse rings to complete graph), where influence propagates iteratively. This allows us to disentangle one-step hub dominance from multi-round interaction. Our contributions are summarized as follows:

\textbf{(1) Topology as a determinant of conformity.}
We show that network structure systematically shapes conformity by trading off convergence speed and robustness. Centralized structures concentrate influence, yielding immediate decisions but making outcomes highly dependent on hub competence and model homogeneity; Distributed structures diffuse influence and support more robust consensus formation, while increased connectivity accelerates convergence and may facilitate high-confidence error cascades.

\textbf{(2) A transparent confidence-normalized pooling rule.} We propose a transparent and generalizable update mechanism that introduces a global parameter to explicitly regulate social influence. By integrating agent-level confidence into the pooling process, the rule yields bounded belief scores, stable binarization, and well-controlled conformity dynamics, providing a principled basis for different decision paradigms.

\textbf{(3) Empirical insights on robustness and risk.} Experiments on current fact-checking datasets reveal that system reliability is heavily contingent on hub competence and majority composition. We further characterize the dual nature of conformity, showing that it enhances accuracy under reliable conditions but can induce confident errors when early majority signals are incorrect.

\section{Related Works}

\subsection{MAS and Collective Decision-Making}

Early research on MAS primarily focused on distributed optimization and consensus formation under idealized assumptions, where agents were modeled as homogeneous entities with limited reasoning capacity \citep{Bao31122022, amirkhani2022consensus}. 
The recent integration of LLMs expands this paradigm by endowing agents with advanced reasoning, enabling MAS to address complex, unstructured decision-making problems such as multi-step problem solving \citep{chen-etal-2024-comm} and automated fact-checking \citep{han2025debatetodetect}.
Despite these advances, most existing studies emphasize aggregate performance metrics, leaving the role of social dynamics such as conformity in shaping collective reliability largely unexamined.

\subsection{Conformity and Opinion Dynamics}
Conformity, defined as the tendency of individuals to align their judgments with a perceived majority, has been widely studied in social psychology \citep{capuano2024systematic}. In computational settings, opinion dynamics models such as DeGroot averaging \citep{dong2024social} and bounded-confidence mechanisms \citep{PhysRevResearch.5.023179} formalize how local interactions give rise to collective consensus and have been used to explain phenomena including information cascades and polarization \citep{shirzadi2025}. However, these models operate at an abstract level and do not capture the semantic reasoning or contextual understanding characteristic of modern LLM agents \citep{aouini2025}. Consequently, how classical conformity theories extend to LLM-based MAS, where judgments are produced through generative processes, remains insufficiently explored.

\subsection{Network Topology}
Network topology is a fundamental determinant of how influence propagates and consensus forms in multi-agent systems \citep{Cheng_Wang_Liu_2021, amirkhani2022consensus}. \textbf{Centralized aggregation} structures, such as star or hierarchical graphs, concentrate influence and enable rapid decision-making, but render collective outcomes highly sensitive to the reliability of central agents. By contrast, \textbf{Distributed consensus} structures, including ring and complete graphs, diffuse influence across agents, enhancing robustness to local noise at the expense of slower convergence. Latest studies have further examined how connectivity modulates communication cost, convergence speed, and robustness \citep{da2025understandinguncertaintyllmexplanations, wang-etal-2025-g, yang2025topologicalstructurelearningresearch}. 
Despite these efforts, existing work has yet to clarify the interplay between interaction topology, agent confidence, and conformity in shaping collective outcomes in LLM-driven MAS.

\section{Methodology}
\label{methodology}
\subsection{Agent Decision Model}

Building on classical opinion dynamics and weighted consensus frameworks \citep{anderson2019recent, PhysRevResearch.5.023179, dong2024social}, we formalize misinformation detection in MAS as a binary collective decision problem. At each interaction round $t$, agent $i$ produces a binary judgment $y_i^{(t)} \in {0,1}$ (with $0$ denoting True and $1$ denoting False), accompanied by a confidence score $p_i^{(t)}$ that quantifies the agent’s self-assessed reliability.

\paragraph{Update rule.}
Agents update an internal support score using confidence-normalized pooling:

\begin{equation}
\label{eq:score}
\small
s_i^{(t+1)} = \frac{\alpha\, p_i^{(t)} y_i^{(t)} + (1-\alpha) \sum_{j \in N_i} p_j^{(t)} y_j^{(t)}}{\alpha\, p_i^{(t)} + (1-\alpha) \sum_{j \in N_i} p_j^{(t)} + \varepsilon},
\end{equation}
where $N_i$ denotes the neighbor set of agent $i$, and $\varepsilon$ ensures numerical stability. The dynamics are governed by two parameters: $\alpha \in [0, 1]$ is a fixed hyperparameter that balances self-reliance and peer influence, and $p_i^{(t)} \in [0,1]$ is a self-reported confidence score generated by LLMs at each round, modulating both the persistence of its judgment and the influence on neighbors. By construction, $s_i^{(t+1)} \in (0, 1)$, while in high self-reliance (large $\alpha$), strong confident neighbor signals can still sway the decision.

\paragraph{Binary readout.}
The score is mapped to a binary label using a fixed threshold $\tau$:
\begin{equation}
y_i^{(t+1)} = \mathbb{1}\big[s_i^{(t+1)} \ge \tau\big],
\label{eq:readout}
\end{equation}
where $\tau = 0.5$ by default, so $s_i^{(t+1)} < 0.5$ yields True ($0$) and $s_i^{(t+1)} \ge 0.5$ yields False ($1$), ensuring class symmetry and stable, interpretable binarization across tasks and confidence distributions.

\subsection{Prompt Design}

Each agent receives a structured prompt comprising three core components:
(i) a concise background profile, automatically generated by LLMs, situating the claim within its domain context;
(ii) a task description instructing the agent to evaluate the claim based on the provided profile and its own reasoning; and
\textbf{(iii) output requirements specifying a binary label $y_i^{(t)} \in \{0,1\}$, a confidence score $p_i \in [0,1]$, and a brief justification}.
These three outputs are then incorporated into the update rule in Eq.~(\ref{eq:score}) to revise the agent’s belief and guide subsequent decisions. Complete prompts and pseudocode are provided in Appendix~\ref{sec:prompt-templates}.

\subsection{Network Topologies}

We analyze conformity dynamics across two distinct topological paradigms: \textbf{Centralized Aggregation} and \textbf{Distributed Consensus}. Centralized structures rely on immediate, hub-mediated consolidation, and distributed architectures foster iterative, emergent consensus. To ensure comparability, all topologies are instantiated with seven fixed agents ($N=7$).
Detailed explanations of the evaluation metrics are provided in Appendix~\ref{sec:formula}.

\subsubsection{Centralized Aggregation}

\textbf{Centralized Aggregation} is characterized by a unidirectional upward information flow from peripheral agents to a central authority, whose synthesis determines the collective outcome. We examine two representative configurations (Figure~\ref{Centralized}):

\begin{figure}[htbp]
\centering
  \includegraphics[width=0.95\linewidth]{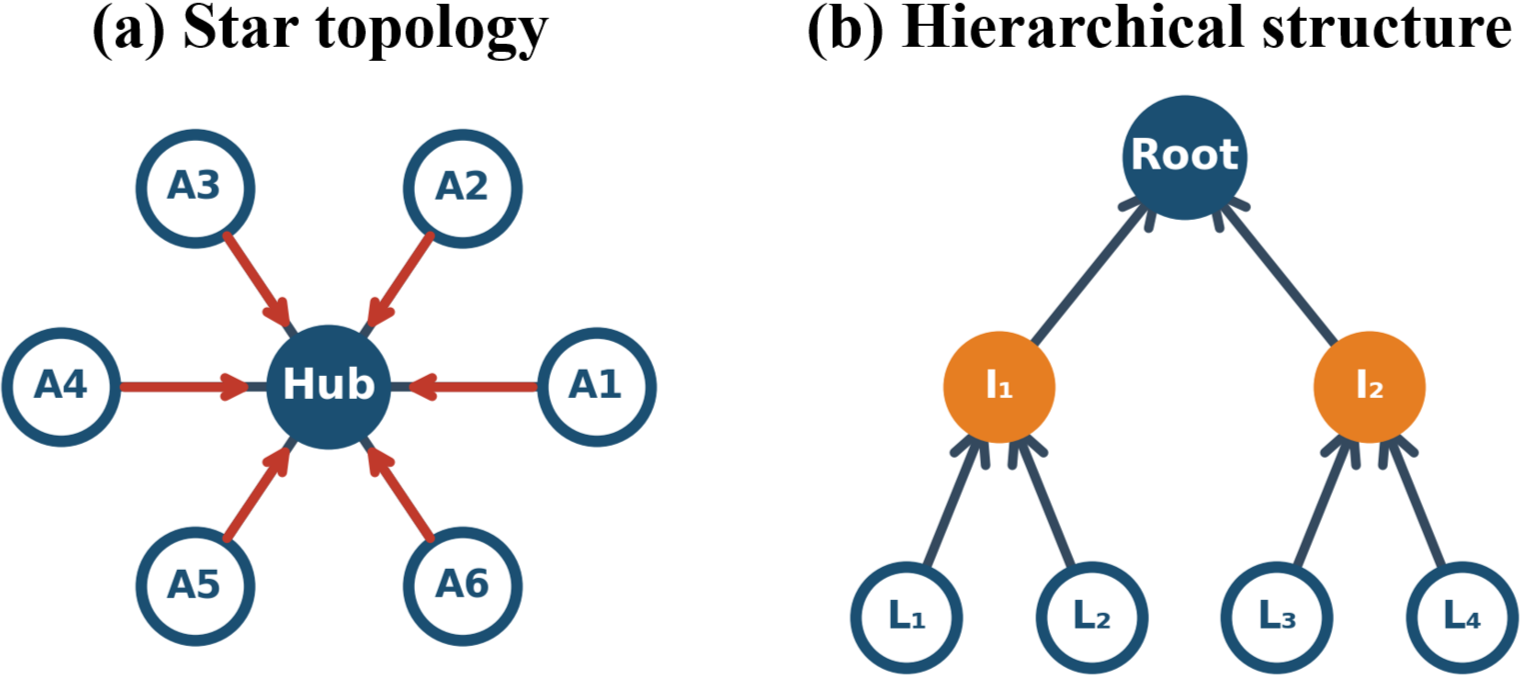}
  \caption{Centralized Aggregation Topology.}
  \label{Centralized}
\end{figure}

\begin{table*}[t]
\centering
\renewcommand{\arraystretch}{0.85}
\begin{tabular}{ccccccccccc}
\toprule
\multirow{2}{*}{\textbf{Topology}} & \multirow{2}{*}{\textbf{$\alpha$}} 
& \multicolumn{3}{c}{\textbf{GPT-3.5}} 
& \multicolumn{3}{c}{\textbf{GPT-4o}} 
& \multicolumn{3}{c}{\textbf{Llama3.3}} \\
\cmidrule(lr){3-5}\cmidrule(lr){6-8}\cmidrule(lr){9-11}
 &  & CA & PA & CPC & CA & PA & CPC & CA & PA & CPC \\
\midrule
\multirow{6}{*}{Star}
 & 0.00 & 0.67 & 0.62 & 0.62 & 0.69 & 0.65 & 0.63 & 0.66 & 0.62 & 0.65 \\
 & 0.25 & 0.73 & 0.68 & 0.73 & 0.75 & 0.71 & 0.77 & 0.74 & 0.72 & 0.68 \\
 & 0.50 & \underline{0.76} & \textbf{0.71} & \underline{0.75} & \underline{0.77} & 0.73 & \underline{0.82} & 0.76 & \underline{0.74} & \underline{0.84} \\
 & 0.75 & \textbf{0.80} & \underline{0.70} & \textbf{0.79} & \textbf{0.82} & \textbf{0.77} & \textbf{0.84} & \textbf{0.80} & \textbf{0.80} & \textbf{0.87} \\
 & 1.00 & 0.75 & 0.69 & 0.74 & \underline{0.77} & \underline{0.75} & \underline{0.75} & \underline{0.78} & \underline{0.74} & 0.73 \\
 & \textit{No-weight} & 0.69 & 0.64 & 0.58 & 0.71 & 0.66 & 0.60 & 0.70 & 0.65 & 0.59 \\
\midrule
\multirow{6}{*}{Hierarchical}
 & 0.00 & 0.66 & 0.63 & 0.67 & 0.68 & 0.67 & 0.67 & 0.70 & 0.65 & 0.61 \\
 & 0.25 & 0.70 & 0.67 & 0.70 & 0.72 & 0.69 & 0.71 & 0.73 & 0.70 & 0.64 \\
 & 0.50 & \underline{0.72} & 0.68 & \textbf{0.77} & \underline{0.76} & 0.70 & \underline{0.80} & \underline{0.77} & 0.71 & \textbf{0.82} \\
 & 0.75 & \textbf{0.75} & \textbf{0.71} & \textbf{0.77} & \textbf{0.78} & \underline{0.73} & \textbf{0.82} & \textbf{0.79} & \textbf{0.75} & \underline{0.80} \\
 & 1.00 & \underline{0.72} & \underline{0.69} & \underline{0.74} & 0.75 & \textbf{0.74} & 0.78 & 0.75 & \underline{0.72} & 0.79 \\
 & \textit{No-weight} & 0.68 & 0.65 & 0.60 & 0.70 & 0.66 & 0.62 & 0.71 & 0.67 & 0.63 \\
\bottomrule
\end{tabular}
\caption{
Centralized Aggregation under the single-round protocol.
Results are reported in terms of central accuracy (CA), peripheral accuracy (PA), and center--periphery consistency (CPC).
Best and second-best results within each topology are highlighted in \textbf{bold} and \underline{underline}, respectively (ties are marked).
The \textit{No-weight} setting removes confidence-weighted aggregation and the global self-weighting parameter.
Across models and topologies, $\alpha = 0.75$ most often achieves the best performance.}

\label{tab:centralized}
\end{table*}

\begin{table*}[ht]
\centering
\renewcommand{\arraystretch}{0.78}
\resizebox{\textwidth}{!}{
\begin{tabular}{cccccccccccccc}
\toprule
\multirow{2}{*}{\textbf{Neighbors}} & \multirow{2}{*}{$\alpha$} 
& \multicolumn{4}{c}{\textbf{GPT-3.5}} 
& \multicolumn{4}{c}{\textbf{GPT-4o}} 
& \multicolumn{4}{c}{\textbf{Llama3.3}} \\
\cmidrule(lr){3-6}\cmidrule(lr){7-10}\cmidrule(lr){11-14}
 &  & FA & TTC & ACI & TT 
    & FA & TTC & ACI & TT
    & FA & TTC & ACI & TT \\
\midrule
\multirow{6}{*}{2} 
 & 0.00 & 0.68 & 7.0 & 0.58 & 9.5k & 0.70 & 6.4 & 0.62 & 9.2k & 0.67 & 6.2 & 0.60 & 7.9k \\
 & 0.25 & 0.70 & 6.5 & 0.61 & 8.9k & 0.73 & 6.2 & 0.67 & 9.0k & 0.69 & 5.9 & 0.66 & 7.5k \\
 & 0.50 & 0.72 & 6.3 & \underline{0.71} & 8.7k & 0.75 & 5.7 & 0.72 & 8.3k & 0.73 & 5.8 & 0.71 & 7.4k \\
 & 0.75 & \textbf{0.75} & \underline{5.8} & \textbf{0.75} & \underline{8.2k} & \textbf{0.79} & \underline{5.1} & \textbf{0.77} & \underline{7.6k} & \textbf{0.78} & \underline{5.6} & \textbf{0.78} & \underline{7.2k} \\
 & 1.00 & \underline{0.74} & \textbf{5.3} & 0.68 & \textbf{7.5k} & \underline{0.78} & \textbf{5.0} & \underline{0.75} & \textbf{7.5k} & \underline{0.76} & \textbf{5.1} & \underline{0.73} & \textbf{6.5k} \\
 & \textit{No-weight} & 0.69 & 7.4 & 0.55 & 9.8k & 0.71 & 6.9 & 0.57 & 9.9k & 0.70 & 6.7 & 0.56 & 8.3k \\
\midrule
\multirow{6}{*}{3} 
 & 0.00 & 0.70 & 7.2 & 0.60 & 10.7k & 0.70 & 6.5 & 0.63 & 10.2k & 0.68 & 6.3 & 0.61 & 8.8k \\
 & 0.25 & 0.72 & 6.5 & 0.64 & 9.8k & 0.74 & 5.8 & 0.70 & 9.3k & 0.70 & 5.9 & 0.69 & 8.0k \\
 & 0.50 & 0.75 & 5.8 & 0.73 & 9.0k & 0.75 & 5.3 & 0.77 & 8.7k & 0.76 & 5.6 & 0.74 & 7.8k \\
 & 0.75 & \textbf{0.78} & \underline{5.1} & \textbf{0.77} & \underline{8.2k} & \textbf{0.82} & \underline{4.7} & \textbf{0.81} & \underline{7.8k} & \textbf{0.79} & \underline{5.0} & \textbf{0.80} & \underline{7.2k} \\
 & 1.00 & \underline{0.77} & \textbf{4.9} & \underline{0.74} & \textbf{8.0k} & \underline{0.80} & \textbf{4.5} & \underline{0.79} & \textbf{7.7k} & \underline{0.78} & \textbf{4.8} & \underline{0.77} & \textbf{7.1k} \\
 & \textit{No-weight} & 0.71 & 6.8 & 0.58 & 10.2k & 0.73 & 6.2 & 0.60 & 9.6k & 0.72 & 6.1 & 0.59 & 8.4k \\
\midrule
\multirow{6}{*}{4} 
 & 0.00 & 0.69 & 6.9 & 0.61 & 11.2k & 0.71 & 6.3 & 0.63 & 11.0k & 0.70 & 6.0 & 0.62 & 9.1k \\
 & 0.25 & 0.71 & 6.1 & 0.69 & 10.1k & 0.74 & 5.4 & 0.73 & 9.6k & 0.72 & 5.7 & 0.73 & 8.4k \\
 & 0.50 & 0.74 & 5.0 & 0.74 & 8.8k & 0.81 & 4.7 & 0.77 & 8.6k & 0.79 & 5.0 & 0.76 & 7.5k \\
 & 0.75 & \textbf{0.78} & \underline{4.3} & \textbf{0.79} & \underline{7.9k} & \textbf{0.83} & \underline{4.1} & \textbf{0.81} & \underline{7.9k} & \textbf{0.81} & \underline{4.2} & \textbf{0.82} & \underline{6.9k} \\
 & 1.00 & \underline{0.77} & \textbf{4.0} & \underline{0.75} & \textbf{7.4k} & \underline{0.82} & \textbf{3.9} & \underline{0.80} & \textbf{7.6k} & \underline{0.80} & \textbf{4.1} & \underline{0.79} & \textbf{6.8k} \\
 & \textit{No-weight} & 0.72 & 6.2 & 0.60 & 10.2k & 0.74 & 5.7 & 0.62 & 9.9k & 0.73 & 5.6 & 0.61 & 8.4k \\
\midrule
\multirow{6}{*}{5} 
 & 0.00 & 0.69 & 6.8 & 0.59 & 11.7k & 0.71 & 6.2 & 0.62 & 11.4k & 0.68 & 6.1 & 0.60 & 9.8k \\
 & 0.25 & 0.72 & 5.5 & 0.70 & 10.0k & 0.78 & 5.3 & 0.73 & 10.3k & 0.74 & 5.1 & 0.74 & 7.9k \\
 & 0.50 & \underline{0.74} & 4.5 & 0.76 & 8.6k & 0.81 & 4.1 & 0.79 & 8.8k & \underline{0.82} & 4.4 & 0.78 & 7.6k \\
 & 0.75 & \textbf{0.75} & \underline{3.7} & \textbf{0.81} & \underline{7.6k} & \textbf{0.84} & \underline{3.5} & \textbf{0.83} & \underline{7.8k} & \textbf{0.84} & \underline{3.4} & \textbf{0.84} & \underline{6.6k} \\
 & 1.00 & \underline{0.74} & \textbf{3.2} & \underline{0.78} & \textbf{6.9k} & \underline{0.83} & \textbf{3.0} & \underline{0.81} & \textbf{7.0k} & 0.81 & \textbf{3.1} & \underline{0.80} & \textbf{6.2k} \\
 & \textit{No-weight} & 0.73 & 5.8 & 0.63 & 10.4k & 0.75 & 5.2 & 0.65 & 10.3k & 0.74 & 5.3 & 0.64 & 8.5k \\
\midrule
\multirow{6}{*}{6} 
 & 0.00 & 0.72 & 6.3 & 0.62 & 11.4k & 0.74 & 5.8 & 0.65 & 11.1k & 0.71 & 5.7 & 0.63 & 9.0k \\
 & 0.25 & 0.75 & 5.0 & 0.74 & 9.4k & 0.77 & 4.6 & 0.77 & 9.3k & 0.78 & 4.7 & 0.78 & 7.3k \\
 & 0.50 & 0.75 & 4.0 & \underline{0.80} & 7.8k & \underline{0.82} & 3.8 & 0.77 & 8.0k & 0.81 & 4.2 & \underline{0.81} & 6.7k \\
 & 0.75 & \textbf{0.78} & \underline{3.8} & \textbf{0.81} & \underline{7.7k} & \textbf{0.85} & \underline{3.4} & \textbf{0.85} & \underline{7.2k} & \textbf{0.83} & \underline{3.2} & \textbf{0.82} & \underline{6.1k} \\
 & 1.00 & \underline{0.77} & \textbf{3.2} & 0.78 & \textbf{6.7k} & 0.80 & \textbf{2.9} & \underline{0.83} & \textbf{6.8k} & \underline{0.82} & \textbf{3.0} & \underline{0.81} & \textbf{5.6k} \\
 & \textit{No-weight} & 0.74 & 5.4 & 0.66 & 10.1k & 0.76 & 5.0 & 0.68 & 9.8k & 0.75 & 5.1 & 0.67 & 7.8k \\
\bottomrule
\end{tabular}}
\caption{
Results of Distributed Consensus under varying neighbor counts $m$ and self-weighting.
Performance is reported in terms of final accuracy (FA), time-to-consensus (TTC), average conformity index (ACI), and total tokens (TT).
For each neighbor-count block, the best and second-best results are highlighted in \textbf{bold} and \underline{underline}, respectively (ties are marked).
The \textit{No-weight} setting removes confidence-weighted aggregation and the global self-weighting parameter. Across models and topologies, $\alpha = 0.75$ most often achieves the best performance.
}
\label{tab:voting}
\end{table*}

\textbf{(a) Star Network:} Six peripheral nodes transmit judgments directly to a central hub without lateral interaction.

\textbf{(b) Hierarchical Structure:} A three-layered tiered architecture where inputs from leaf nodes are aggregated by intermediate agents before reaching the root.

\paragraph{Protocol.} In both structures, decision-making is executed in a single update round. Peripheral or lower-tier agents submit judgments to the hub or root, which integrates these inputs to generate a final output. This central output is adopted as the group's collective decision.

\paragraph{Metrics.} We assess system reliability and hub influence with three metrics:
\textbf{(1) Central Accuracy (CA):} Correctness of the hub/root node's final judgment against ground truth. 
\textbf{(2) Peripheral Accuracy (PA):} Mean correctness of all non-central nodes. 
\textbf{(3) Center–Periphery Consistency (CPC):} The proportion of peripheral nodes aligned with the central decision, quantifying immediate conformity intensity.

\subsubsection{Distributed Consensus}

\textbf{Distributed Consensus} distributes influence symmetrically, preventing any single node from dominating the decision process. We explore a connectivity spectrum ranging from sparse rings to complete graphs (Figure~\ref{Distributed1}):

\begin{figure}[htbp]
\centering
  \includegraphics[width=0.95\linewidth]{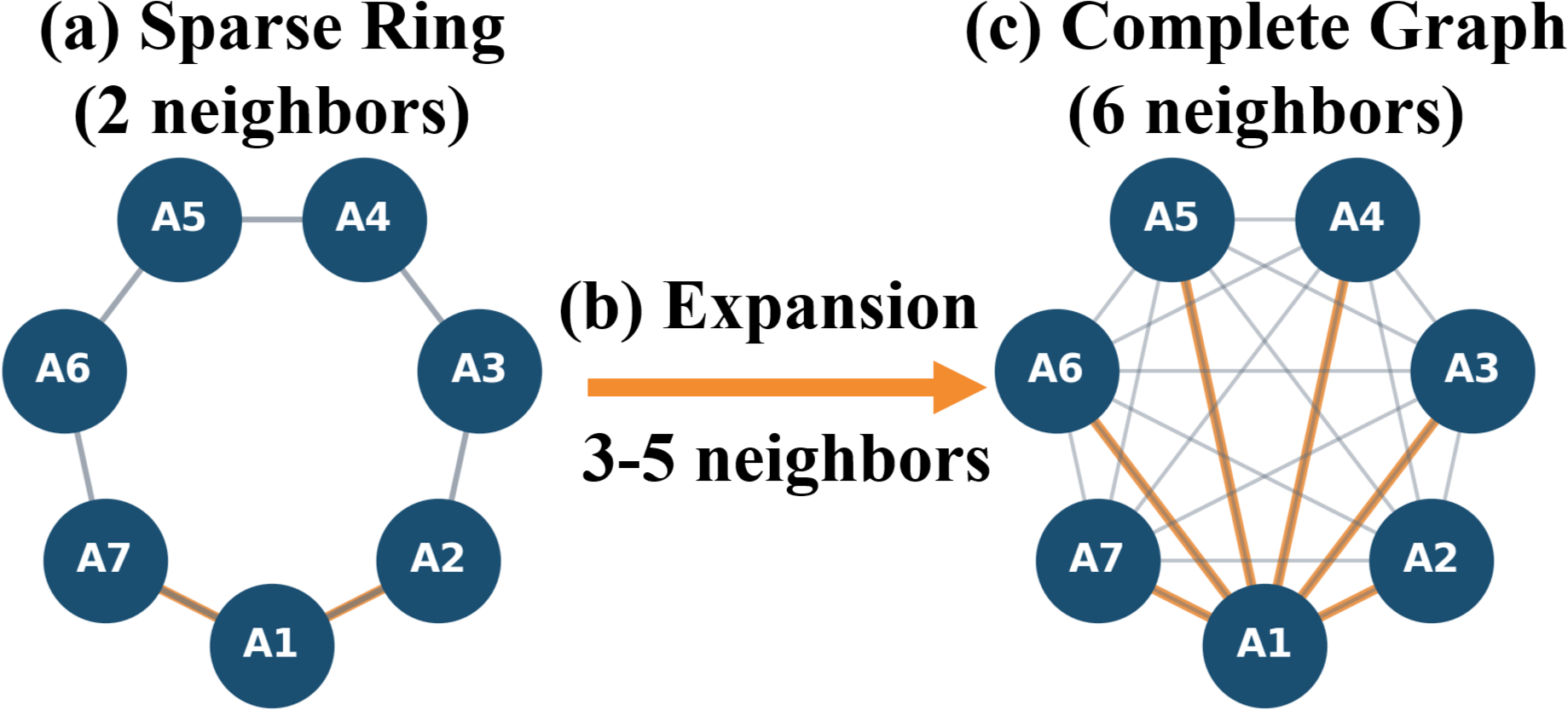}
  \caption{Distributed Consensus Topology.}
  \label{Distributed1}
\end{figure}

\textbf{(a) Sparse ring (2 neighbors)}: Agents interact strictly with their immediate predecessor and successor.

\textbf{(b) Expanded rings (3--5 neighbors)}: Agents connect to a broader local vicinity, accelerating information diffusion.

\textbf{(c) Complete graph (6 neighbors)}: Agents connect to all others, maximizing potential conformity intensity.

\paragraph{Protocol.}
Decision-making operates via iterative rounds. In each step, agents exchange states with neighbors and update their internal confidence and judgment according to Eq.~(\ref{eq:score}). The process terminates upon full consensus (all agents hold the same judgment) or when the maximum round limit $T_{\max}$ is reached.

\begin{figure*}[htbp]
\centering
  \includegraphics[width=0.9\linewidth]{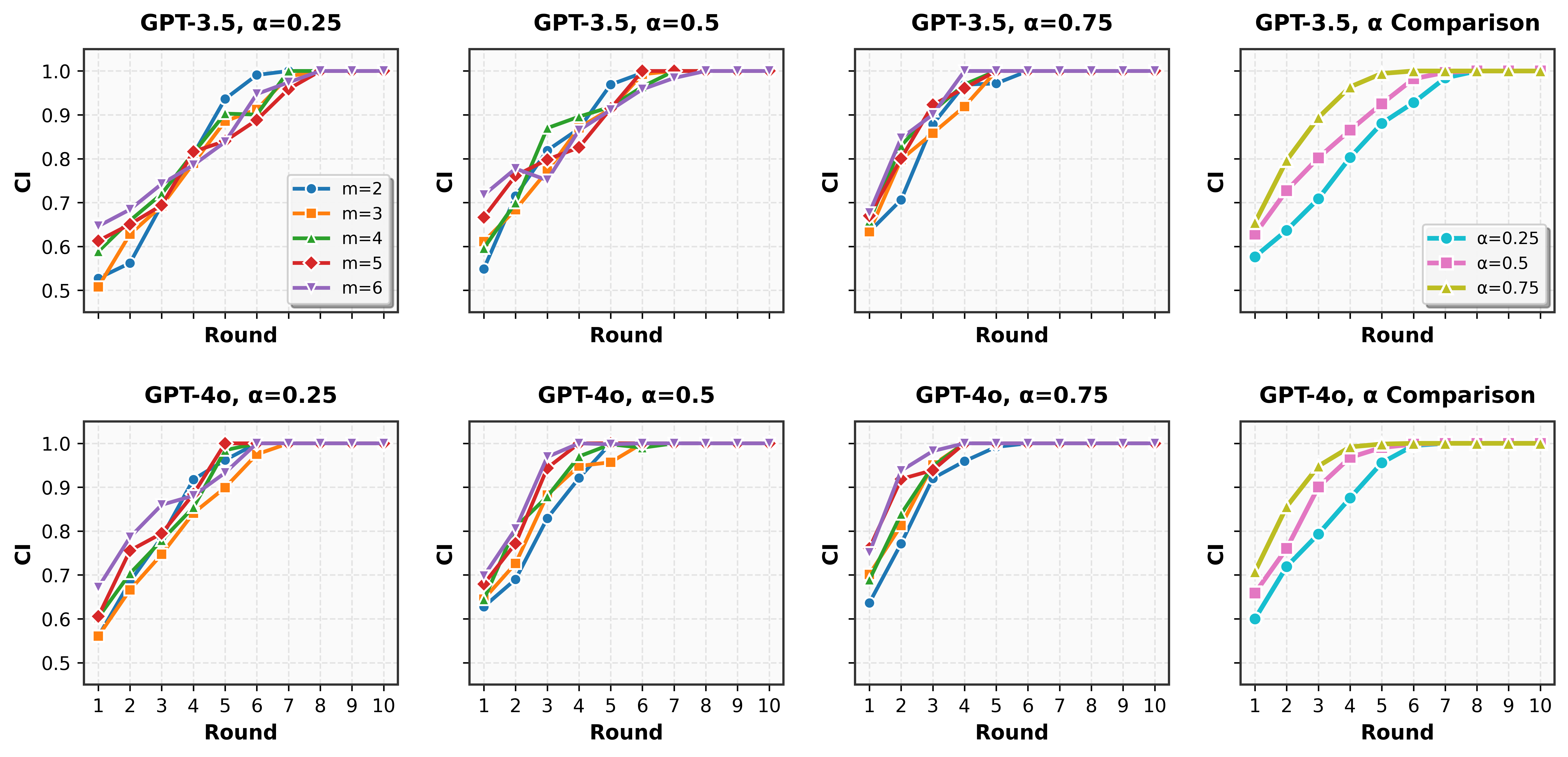}
  \caption{Temporal evolution of the Conformity Index across iterative rounds under varying network densities and self–social weighting, revealing rapid early alignment and diminishing marginal gains in dense structures.}
  \label{fig:ci-trajectories}
\end{figure*}

\paragraph{Metrics.}  
We evaluate the dynamics of emergent consensus with four metrics:
\textbf{(1) Final Accuracy (FA)}: Concordance of the group decision (unanimous or majority view at $T_{\max}$) with the ground truth.
\textbf{(2) Time-to-Consensus (TTC)}: The number of rounds required to reach unanimity. It is undefined if the system fails to converge within $T_{\max}$.
\textbf{(3) Conformity Index (CI)}: Degree of within-group agreement at round $t$, defined as the proportion of agents adopting the majority label. Since judgments are binary, CI is bounded in $[0.5,1]$, where CI $=1$ indicates unanimity and CI $=0.5$ corresponds to an evenly split group. We report \textbf{Average CI (ACI)} as the mean CI over rounds up to $T_{\max}$.
\textbf{(4) Total Tokens (TT)}: The cumulative cost, summing token usage across all agent outputs and interaction rounds.

\section{Experiment}
\label{experiment}
\subsection{Experimental Setup}

\paragraph{Dataset.} We collect Snopes25, a new benchmark comprising 448 real-world claims (252 false, 196 true) fact-checked by Snopes editors. All claims are from January to June 2025 to minimize potential data contamination from pre-trained knowledge.

\paragraph{Implementation.} 
Experiments are conducted using two proprietary models, GPT-3.5~\citep{openai2023gpt35turbo} and GPT-4o~\citep{openai2024hello}, and one open-source model, Llama3.3-70B-Instruct~\citep{llama3modelcard} to ensure all models’ training cutoff precede our evaluation data. Detailed hyperparameters and zero-shot baselines are provided in Appendix~\ref{sec:model-settings}. For \textbf{Centralized Aggregation}, decision-making is restricted to a single update round, capturing immediate conformity to the hub or root node. For \textbf{Distributed Consensus}, agents iteratively exchange judgments for up to $T_{\max}=10$ rounds or until consensus is reached.
% Agent’s individual confidence $p_i$ is generated by the LLM, while the self-weighting parameter $\alpha$ is fixed in each run. 

\paragraph{Parameter Configurations.}
We consider five values of $\alpha$: 0 (fully conformist),
0.25 (socially influenced), 0.50 (balanced), 0.75 (self-reliant),
and 1 (fully independent), capturing different predispositions toward
peer influence versus self-reliance. All reported metrics represent the mean of 10 independent runs, with standard errors consistent at $<0.09$. To isolate the contribution of our confidence-weighting mechanism, we introduce a \textit{No-weight} baseline that removes both the self-weighting parameter $\alpha$ and the confidence score $p$, forcing agent updates to depend solely on the textual context of neighbor interactions and isolating the effect of explicit confidence weighting.

\subsection{Results on Centralized Aggregation}

We first evaluate \textbf{Centralized Aggregation}, and the results are summarized in Table~\ref{tab:centralized}.
Across both topologies and all models, central accuracy (CA) increases monotonically with the self-weighting parameter $\alpha$, indicating that stronger self-reliance enables the hub to better filter peripheral noise. For example, in the star topology, CA for GPT-3.5 improves from 0.67 at $\alpha=0$ to 0.80 at $\alpha=0.75$. Hierarchical structures follow the same trend, though with attenuated improvements due to information dilution across levels. 

In contrast, peripheral accuracy (PA) remains largely stable across $\alpha$, indicating that variations in self–social weighting primarily influence the quality of the hub’s aggregation rather than improving local agent correctness. Center–periphery consistency (CPC) increases with $\alpha$ across models and topologies, suggesting that as the hub places greater emphasis on its own confidence-weighted judgment, its final decision becomes more stable and, given sufficient hub competence, more likely to align with peripheral votes.

\subsection{Results on Distributed Consensus}

We next turn to \textbf{Distributed Consensus}, and Table~\ref{tab:voting} summarizes performance across varying network densities and self-social weighting. Final Accuracy (FA) generally improves with connectivity $m$ and often peaks at moderate-to-high self-reliance ($\alpha=0.75$). For instance, with GPT-3.5 under a socially influenced setting ($\alpha=0.25$), increasing connectivity from sparse rings ($m=2$) to complete graphs ($m=6$) raises accuracy from 0.70 to 0.75. Denser networks facilitate rapid information propagation and thus accelerate convergence, whereas decreasing $\alpha$ tends to strengthen conformity pressure and can further speed up consensus, albeit at the risk of amplifying early biases. In sparse networks ($m=2$) with strong peer influence ($\alpha=0.25$), conformity remains moderate, whereas in highly connected regimes ($m=6$), groups exhibit high homogeneity.

Without the confidence-weighted mechanism, FA degrades uniformly across all topologies, and simply increasing connectivity fails to recover the gains of the full model. While increasing neighbors inflates the volume of messages per round, higher $m$ substantially reduces the number of rounds required to reach consensus. In summary, sparse networks preserve diversity and mitigate premature convergence, while complete graphs accelerate consensus but may increase susceptibility to information cascades when early signals are biased.

\begin{table*}[ht]
\centering
% \small
\renewcommand{\arraystretch}{0.8}
\resizebox{\textwidth}{!}{
\begin{tabular}{cccccccccc}
\toprule
\multirow{2}{*}{\textbf{Setting}} & \multicolumn{3}{c}{\textbf{Model Assignment}} & \multicolumn{5}{c}{\textbf{Performance}} \\
\cmidrule(lr){2-4} \cmidrule(lr){5-9}
& Hub & Left & Right & CA & PA$_L$ & PA$_R$ & CPC$_L$ & CPC$_R$ \\
\midrule
Cap-H1 & GPT-4o        & GPT-4o        & GPT-3.5          & 0.82 & 0.79 & 0.67 & 0.87 & 0.74 \\
Cap-H2 & GPT-3.5       & GPT-3.5       & GPT-4o           & 0.77 & 0.73 & 0.77 & 0.81 & 0.69 \\
Cap-H3 & Llama3.3-70B  & Llama3.3-70B  & Llama3.3-8B      & 0.80 & 0.76 & 0.70 & 0.86 & 0.78 \\
Cap-H4 & Llama3.3-8B   & Llama3.3-8B   & Llama3.3-70B     & 0.76 & 0.71 & 0.75 & 0.80 & 0.72 \\
Type-H1 & GPT-4o       & GPT-4o        & Llama3.3-70B         & 0.82 & 0.75 & 0.73 & 0.88 & 0.76 \\
Type-H2 & Llama3.3-70B     & Llama3.3-70B      & GPT-4o           & 0.79 & 0.72 & 0.76 & 0.84 & 0.77 \\
\bottomrule
\end{tabular}}
\caption{Heterogeneous Centralized Aggregation. The Left branch is aligned with the hub (same backbone), while the Right branch is assigned a different model. We report Central Accuracy (CA), branch-level Peripheral Accuracy (PA$_L$, PA$_R$), and Center--Periphery Consistency (CPC$_L$, CPC$_R$). Centralized performance is primarily driven by hub competence rather than peripheral model strength.}
\label{tab:hetero}
\end{table*}

\begin{figure*}[htbp]
\centering
  \includegraphics[width=0.78\linewidth]{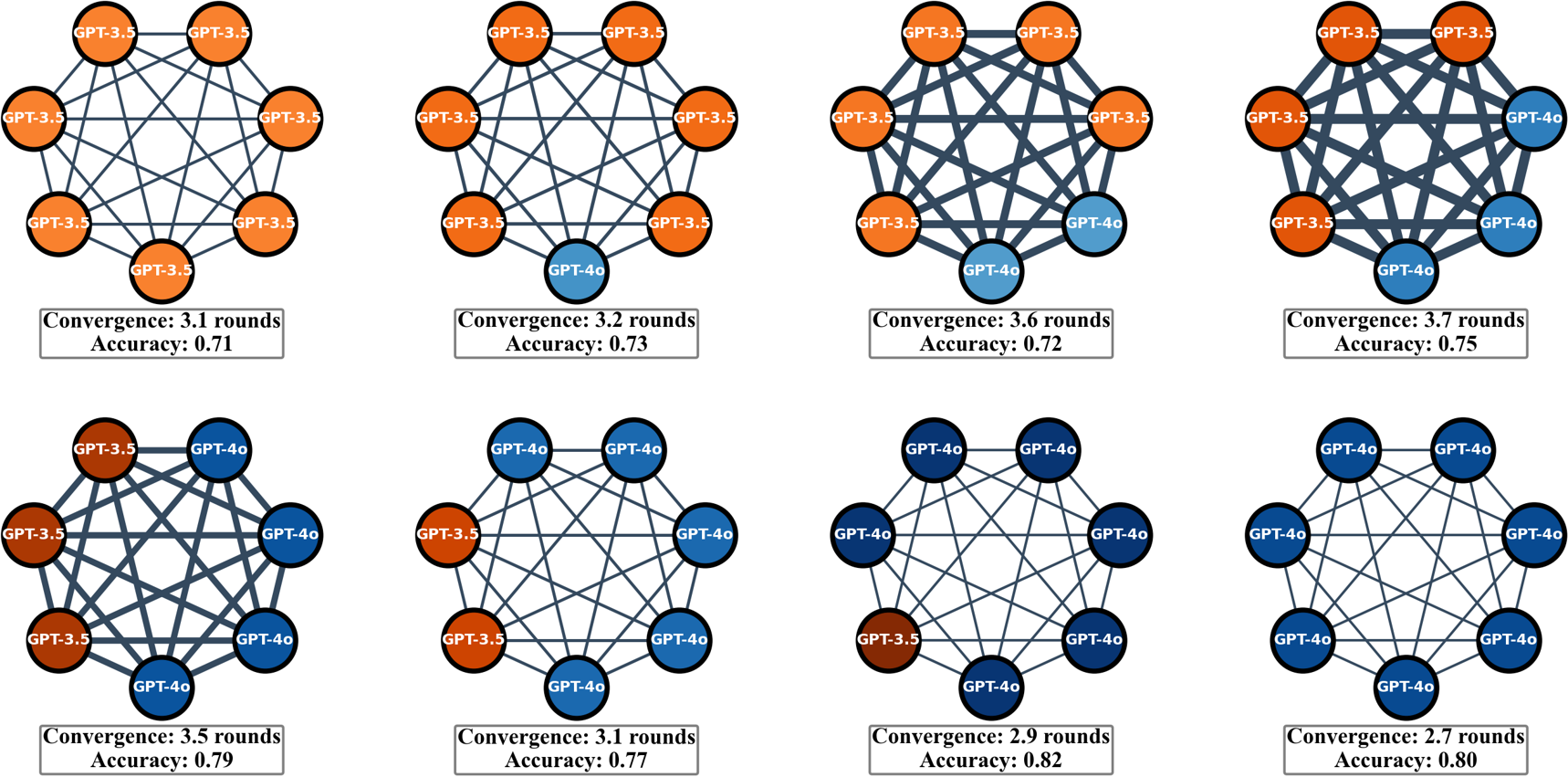}
  \caption{Distributed Consensus with heterogeneous model composition. Darker node colors indicate higher accuracy, while thicker edges signify prolonged time-to-consensus. Both the proportion and structure of stronger agents jointly influence convergence speed and final collective accuracy.}
  \label{fig:11}
\end{figure*}

Figure~\ref{fig:ci-trajectories} further illustrates the temporal evolution of the Conformity Index (CI) across rounds under different neighbor counts $m$ and self--social weighting. CI rises markedly faster in denser networks, indicating that quicker consolidation of group alignment once a dominant opinion emerges. Specifically, sparse rings ($m=2$) typically start from 0.55-0.65 and may take six or more rounds to approach unanimity, whereas denser structures ($m\geq 4$) exceed 0.75 by the second round and nearly converge by the fourth. Across all settings, the steepest CI increase occurs within the first four rounds, followed by a slower consolidation phase until consensus, and the gains from increasing $\alpha$ or enlarging $m$ exhibit diminishing returns once networks are sufficiently dense. 

\begin{figure*}[htbp]
\centering
  \includegraphics[width=0.9\linewidth]{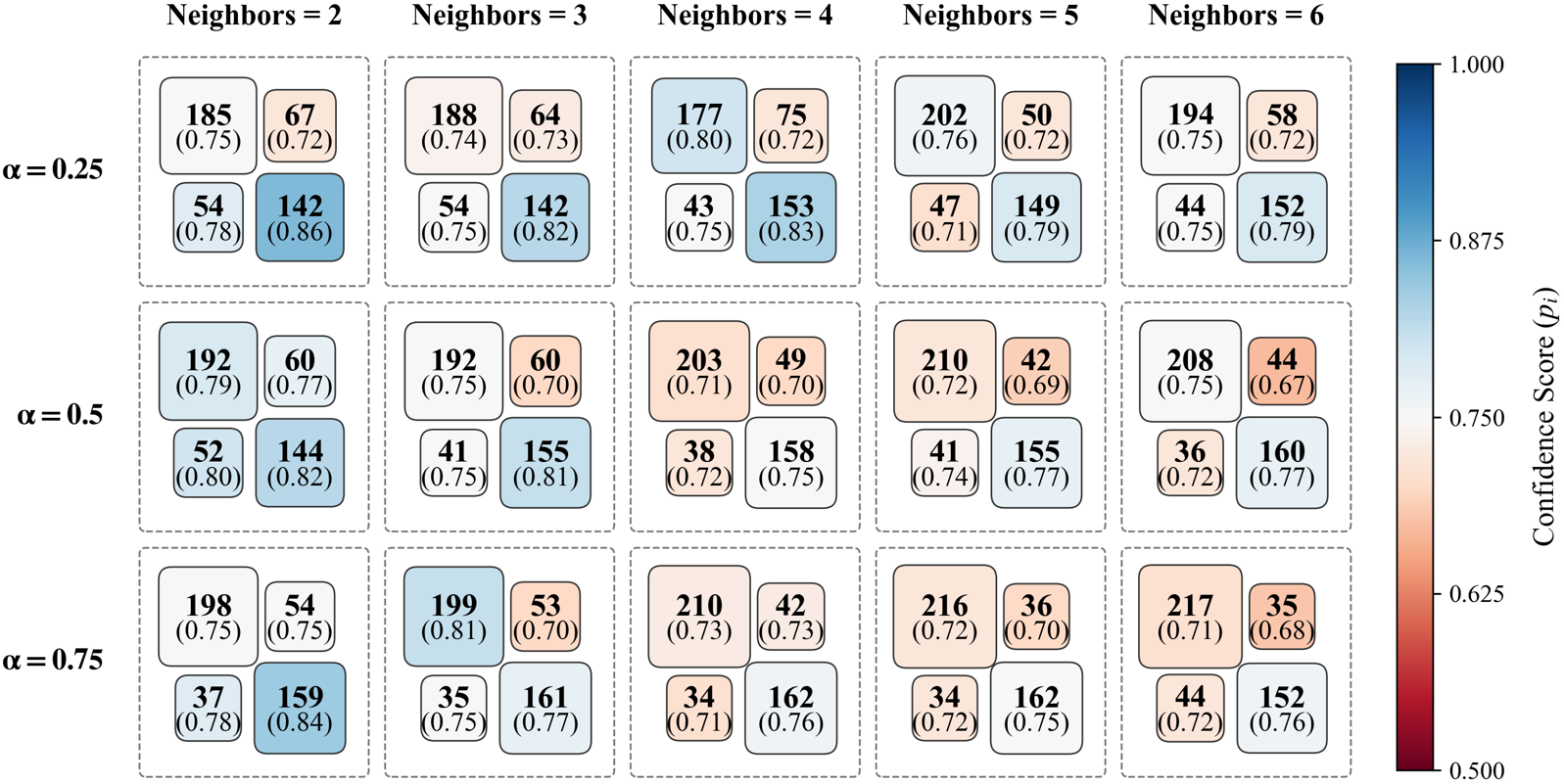}
  \caption{Confusion-matrix heatmaps of GPT-4o under \textbf{Distributed Consensus}, varying neighbor counts and $\alpha$ values. Each cell reports group-level outcomes: the number of correctly classified true claims (top-left), correctly classified false claims (top-right), misclassified true claims (bottom-left), and misclassified false claims (bottom-right). Cell size reflects the frequency, while color encodes the mean confidence score $p_i$. Higher connectivity and lower self-reliance produce more high-confidence misclassifications, highlighting a key failure mode of conformity.}
  \label{fig:confusion}
\end{figure*}

\section{Discussion: Factors Influencing the Conformity in MAS}

\subsection{Model Heterogeneity}

To characterize model heterogeneity in LLM-based MAS, agents are instantiated with different LLM backbones. In \textbf{Centralized Aggregation}, the hub shares its backbone with the Left branch (the \emph{Aligned Branch}), while the Right branch is instantiated with a different model (the \emph{Opposing Branch}). We fix the self--social weighting at $\alpha=0.75$ and report hub-level Central Accuracy (CA), branch-level Peripheral Accuracy (PA$_L$, PA$_R$), and Center--Periphery Consistency (CPC$_L$, CPC$_R$) over 10 independent runs per claim.

Table~\ref{tab:hetero} reveals two salient phenomena associated with centralized conformity. 
First, a consistent \emph{same-model alignment bias} emerges: the hub aligns more frequently with the Aligned Branch than with the Opposing Branch, evidenced by CPC$_L$ being consistently higher than CPC$_R$ across all settings. This pattern suggests that the hub is more receptive to peer rationales that resemble its own reasoning style and inductive biases \citep{doi:10.1073/pnas.2415697122}.
Second, system reliability is predominantly capped by the hub's competence: CA is higher when the hub is instantiated with a stronger model, even when the periphery contains a stronger model on the Opposing Branch. Together, these results highlight a vulnerability of centralized MAS: peripheral diversity does not reliably compensate for central incompetence; instead, the hub can act as a selective filter that amplifies aligned inputs while down-weighting divergent perspectives.

We further examine heterogeneity in \textbf{Distributed Consensus} by varying the ratio of GPT-3.5 to GPT-4o ($0{:}7$ to $7{:}0$) in complete graphs. Experiments are conducted on 50 claims with 10 runs per claim at $\alpha=0.75$. As illustrated in Figure~\ref{fig:11}, heterogeneous groups converge more slowly than homogeneous ones, consistent with increased deliberative friction induced by divergent reasoning priors and confidence calibration. In contrast, homogeneous groups converge more rapidly, and groups in which GPT-4o agents constitute the majority consistently achieve higher final accuracy. This indicates that while homogeneity primarily governs convergence speed, the accuracy of the emergent consensus is ultimately determined by the capability of the prevailing model class.

\subsection{Double-Edged Effect of Conformity}

When biased signals dominate, conformity can amplify individual errors into systematic collective failure \citep{han2025debatetodetect}. We examine these failure modes in GPT-4o under \textbf{Distributed Consensus} through the confusion-matrix heatmaps (Figure~\ref{fig:confusion}). First, increased connectivity exposes agents to a broader set of peer signals, which attenuates individual evidence while promoting group-level alignment. Second, larger $\alpha$ (stronger self-reliance) generally improves accuracy, confirming that maintaining sufficient independence is essential for resisting misleading social influence. However, groups occasionally converge on false claims with high certainty, entrenching misinformation rather than correcting it. Such patterns reveal the ambivalent nature of conformity: it enhances reliability under balanced conditions, yet can solidify collective hallucinations when early biases dominate.

\section{Conclusion}

Our study shows that conformity in LLM-based MAS is shaped by both network topology and self-social weighting. In Centralized Aggregation, the reliability of collective outcomes is tightly coupled to hub competence, with stronger hubs amplifying overall system accuracy. In Distributed Consensus, conformity arises from iterative local interactions, with denser connectivity simultaneously enhancing convergence efficiency and average accuracy while heightening susceptibility to information cascades. While conformity facilitates consensus, it risks cementing high-confidence errors when initial biases prevail. Consequently, practical MAS design must balance efficiency and robustness through careful topology design, weighting calibration, and control of premature convergence.

\section*{Limitations}

\textbf{Modeling Assumptions.} 
First, the proposed confidence-normalized pooling rule relies on self-reported confidence generated by LLMs, which is not guaranteed to be well calibrated and may partially reflect stylistic assertiveness rather than a faithful probability of correctness. Moreover, the self--social weighting parameter is fixed per run and shared across agents, and most distributed-consensus experiments restrict interaction to a fixed, structured per-round update template rather than open-ended deliberation, questioning, or tool-augmented evidence seeking. 

\textbf{Scale and Model Selection.} 
Second, all experiments are conducted with a fixed group size ($N=7$) and a limited set of three LLMs. This design is motivated by three considerations. First, many contemporary MAS deployments remain modest in size due to coordination and cost constraints. Second, fixing $N=7$ enables strict alignment with the three-level hierarchical configuration ($1$ root, $2$ intermediate aggregators, and $4$ leaves). Third, the selected models are chosen primarily to avoid knowledge leakage arising from recent knowledge cutoffs. Our objective is not to maximize the detection accuracy, but to examine how agent-level interaction mechanisms shape the collective behavior. As a result, while stronger models may shift overall performance levels, we expect the relative interaction effects identified in this study to remain qualitatively robust.

\textbf{Task and Domain Scope.} 
Third, the study is restricted to binary misinformation detection. While this task is well suited for isolating conformity dynamics, it represents only a narrow class of collective reasoning problems. More open-ended tasks, such as multi-hop reasoning, policy analysis, or creative collaboration, may exhibit qualitatively different interaction patterns in which conformity interacts with exploration, role specialization, or strategic behavior beyond the scope of the current framework.

\bibliography{iclr2026_conference}

\appendix

% \begin{figure*}[htbp]
% \centering
% \tcolorbox[colback=green!5!white, colframe=green!50!black,
%           title=Example Prompt, fonttitle=\bfseries,
%           coltitle=black, boxrule=0.5pt, arc=2mm, width=\textwidth]
% \textbf{Claim:} Pathogens can be released into the air when a toilet is flushed without a closed lid.  
% \tcblower
% \textbf{Background Profile:} You are an expert in public health specializing in how everyday human activities influence the transmission of infectious diseases...

% \textbf{Task Description:} Carefully read the claim above and evaluate its veracity. Your task is to: 

% 1. Decide whether the claim is True or False.  

% 2. Provide a confidence score $p$ in [0,1] reflecting how certain you are. 

% 3. Give a justification (less than 100 words) explaining the reasoning behind your judgment.  

% 4. Return your answer strictly in the JSON format.
% \endtcolorbox
% \caption{Example Prompt for the Misinformation Detection Task.}
% \label{fig:example-prompts}
% \end{figure*}

\section{Prompt Templates}
\label{sec:prompt-templates}

\subsection{Centralized Aggregation}

\subsubsection{Peripheral Agent (Leaf): One-shot Judgment}
\noindent\textit{Use:} Leaf/peripheral agent issues a single-round decision; no peer inputs are seen.

\paragraph{Prompt.}
\begin{PromptBlock}
You are a careful fact-checking agent. Read the claim and produce a binary
decision and a calibrated confidence.

[Task]
1) Decide whether the claim is True (=0) or False (=1).
2) Provide a confidence p in [0, 1].
3) Give a concise justification (<100 words) grounded in the background profile
  and logical reasoning. Avoid speculation.

[Background Profile]
{{BACKGROUND}}

[Claim]
{{CLAIM}}

[Output Requirements]
Output strictly the following JSON object:
{
  "y": 0 or 1,      // 0=True, 1=False
  "p": number,      // float in [0,1], with up to 2 decimals
  "just": "..."     // <100 words, no markdown
}
Do NOT include any extra keys, prose, or formatting.
\end{PromptBlock}

\subsubsection{Hub/Root: One-shot Judgment with Submitted Leaves}
\noindent\textit{Use:} The hub reads the claim and sees peer JSON reports; weighting is handled by Eq.~(1).

\paragraph{Prompt.}
\begin{PromptBlock}
You are the central (hub/root) fact-checking agent. You will read the claim
and also see a list of peer reports. Produce your own final judgment.

[Claim]
{{CLAIM}}

[Peer Reports]
List of JSON objects from peripheral agents:
{{PEER_JSON_LIST}}
/*
Each element is:
{"agent_id": "Li_i", "y": 0|1, "p": [0,1], "just": "..."}
Do not copy them verbatim in your output. Use them only as additional evidence.
*/

[Task]
1) Decide whether the claim is True (=0) or False (=1).
2) Provide a calibrated confidence p in [0, 1].
3) Give a concise justification (<100 words) that references the most
  diagnostic considerations. Avoid majority-following; reason on merits.

[Output Requirements]
Output strictly the following JSON object:
{
  "y": 0 or 1,
  "p": number,
  "just": "..."
}
Do NOT include any extra keys, prose, or formatting.
\end{PromptBlock}

\subsection{Distributed Consensus}

\subsubsection{Initial Judgment}
\noindent\textit{Use:} Each agent produces its initial decision before any interaction.

\paragraph{Prompt.}
\begin{PromptBlock}
You are an autonomous agent in a distributed consensus setting (no central
controller). Produce your initial judgment.

[Background Profile]
{{BACKGROUND}}

[Claim]
{{CLAIM}}

[Task]
1) Output y in {0,1} where 0=True and 1=False.
2) Output a calibrated confidence p in [0,1].
3) Provide a brief justification (<100 words) grounded in facts and logic.

[Output Requirements]
Output strictly the following JSON object:
{
  "agent_id": "{{AGENT_ID}}",
  "t": 0,
  "y": 0 or 1,
  "p": number,
  "just": "..."
}
Do NOT include any extra keys, prose, or formatting.
\end{PromptBlock}

\subsubsection{Interactive Rounds}
\noindent\textit{Use:} At each round $t\ge 1$, the agent receives its previous state and neighbor reports.

\paragraph{Prompt.}
\begin{PromptBlock}
You are participating in a multi-round distributed consensus process. You will
see your previous stance and your neighbors' reports at round t-1. Provide your
current stance, but note that the final state is computed by the system.

[Context]
• Agent: {{AGENT_ID}}
• Round: {{ROUND}}
• Your previous state (t-1): {"y": {{Y_PREV}}, "p": {{P_PREV}}}
• Neighbor reports (t-1): {{NEIGHBOR_JSON_LIST}}

[Task]
1) State your CURRENT stance (y in {0,1}, p in [0,1]).
2) Provide a compact justification (<80 words) referencing decisive signals.
3) Do not summarize all neighbors; mention only what materially changes your stance.

[Output Requirements]
Output strictly the following JSON object:
{
  "agent_id": "{{AGENT_ID}}",
  "t": {{ROUND}},
  "y": 0 or 1,
  "p": number,
  "just": "..."
}
Do NOT include any extra keys, prose, or formatting.
\end{PromptBlock}

\subsection{Pseudocode for Agent Decision Process}
\label{appendix:pseudocode}

The following pseudocode describes the agent's decision-making process and belief update mechanism:

\begin{lstlisting}[language=Python, caption={Pseudocode for the Agent Decision Process}]
# Initialize agent's state (beliefs, confidence)
agents = initialize_agents()

for t in range(T):  # Loop for T rounds of decision-making and belief updates
    for agent in agents:
        # Step 1: Evaluate claim using the prompt (Figure 2)
        claim = get_claim(t)
        background_profile = generate_background_profile(claim)
        task_description = generate_task_description()
        
        # Agent produces a binary judgment and confidence score
        y_i_t, p_i_t = evaluate_claim(agent, background_profile, task_description)
        
        # Step 2: Update agent's belief score using the update rule (Equation 1)
        s_i_t_plus_1 = update_belief(agents, agent, y_i_t, p_i_t)
        
        # Step 3: Assign updated belief back to the agent
        agent.belief_score = s_i_t_plus_1

# End of simulation
\end{lstlisting}

\section{Model Settings}
\label{sec:model-settings}

For the closed-source LLMs (GPT-3.5 and GPT-4o), we set the sampling temperature to $0.7$ for all runs. 
We use \texttt{gpt-3.5-turbo-16k-0613} and \texttt{gpt-4o-1120}. 
For the open-source model \texttt{llama3.3-70b-instruct}, we adopt the same temperature to ensure comparability; all other decoding hyperparameters follow the default configuration of Ollama.\footnote{\url{https://github.com/ollama/ollama}} We also report the zero-shot performance of each model in Table~\ref{tab:zeroshot}.

\begin{table}[t]
\centering
\setlength{\tabcolsep}{5pt}
\caption{Zero-shot performance of LLMs on Snopes25.}
\label{tab:zeroshot}
\begin{tabular}{lcccc}
\toprule
\textbf{Model} & \textbf{Acc.} & \textbf{Prec.} & \textbf{Rec.} & \textbf{F1} \\
\midrule
GPT-3.5   & 0.64 & 0.62 & 0.60 & 0.61 \\
GPT-4o    & 0.70 & 0.66 & 0.69 & 0.67 \\
Llama3.3  & 0.67 & 0.63 & 0.66 & 0.65 \\
\bottomrule
\end{tabular}
\end{table}

\section{Cases}
\label{sec:cases}

We present representative case studies from the Snopes25 dataset to illustrate
both successful and failure modes of our system.
Cases C.1--C.4 demonstrate correct collective judgments under different
topologies, while Case C.5 highlights a failure scenario in which distributed
consensus amplifies an initially incorrect interpretation.

\subsection{Hierarchical Centralized Aggregation (Correct)}

\paragraph{Claim.}
\emph{Pathogens can be released into the air when a toilet is flushed without a closed lid.}

\paragraph{Ground Truth.}
True.

\paragraph{Topology and Protocol.}
A three-level hierarchy with seven agents: a root (R), two intermediate
aggregators (M$_L$, M$_R$), and four leaves (L1--L4).
A single-round protocol is used: leaves issue one-shot judgments; intermediates
summarize assigned leaves; the root produces the final decision.

\paragraph{Leaf Outputs ($t=0$).}
\begin{PromptBlock}
{"agent_id":"L1","t":0,"y":0,"p":0.71,"just":"Evidence on toilet plume aerosolization indicates airborne release without a closed lid."}
{"agent_id":"L2","t":0,"y":0,"p":0.68,"just":"Mechanistic fluid dynamics and observed droplet formation support potential airborne dispersion."}
{"agent_id":"L3","t":0,"y":0,"p":0.76,"just":"Studies show particle counts rise after flushing; lids reduce but absence increases emission."}
{"agent_id":"L4","t":0,"y":0,"p":0.73,"just":"Reported bioaerosols align with the claim under open-lid flushing."}
\end{PromptBlock}

\paragraph{Intermediate Aggregators.}
\begin{PromptBlock}
{"y":0,"p":0.82,"just":"Leaf reports consistently indicate aerosolized particles after flushing without a lid."}
{"y":0,"p":0.80,"just":"Multiple leaves cite increased particle counts and bioaerosols; the claim is supported."}
\end{PromptBlock}

\paragraph{Root Decision.}
\begin{PromptBlock}
{"y":0,"p":0.86,"just":"Both sub-aggregators converge on airborne release under open-lid flushing."}
\end{PromptBlock}

% -----------------------------
\subsection{Star Centralized Aggregation (Correct)}

\paragraph{Claim.}
\emph{Pathogens can be released into the air when a toilet is flushed without a closed lid.}

\paragraph{Ground Truth.}
True.

\paragraph{Topology and Protocol.}
A star topology with one hub (H) and six leaves (L1--L6).
Leaves issue one-shot judgments; the hub produces the final decision.

\paragraph{Leaf Outputs ($t=0$).}
\begin{PromptBlock}
{"agent_id":"L1","t":0,"y":0,"p":0.72,"just":"Open-lid flushing generates toilet plumes with aerosolized particles."}
{"agent_id":"L2","t":0,"y":0,"p":0.69,"just":"Droplet and aerosol formation supports airborne release without a lid."}
{"agent_id":"L3","t":0,"y":0,"p":0.75,"just":"Particle counts increase after flushing; lids mitigate emissions."}
{"agent_id":"L4","t":0,"y":0,"p":0.71,"just":"Bioaerosol evidence aligns with open-lid flushing."}
{"agent_id":"L5","t":0,"y":0,"p":0.74,"just":"Fluid dynamics indicate an upward plume capable of suspending microbes."}
{"agent_id":"L6","t":0,"y":0,"p":0.70,"just":"Observed plume height supports airborne dispersal."}
\end{PromptBlock}

\paragraph{Hub Decision.}
\begin{PromptBlock}
{"y":0,"p":0.86,"just":"All leaves converge on plume and aerosol evidence; the claim is true."}
\end{PromptBlock}

% -----------------------------
\subsection{Ring Distributed Consensus (2 Neighbors, Correct)}

\paragraph{Claim.}
\emph{Pathogens can be released into the air when a toilet is flushed without a closed lid.}

\paragraph{Ground Truth.}
True.

\paragraph{Topology and Outcome.}
A ring of seven agents, each connected to two neighbors.
The system reaches unanimous consensus ($y=0$) at round $t=3$.

\paragraph{Initial Judgments ($t=0$).}
\begin{PromptBlock}
{"agent_id":"A1","t":0,"y":0,"p":0.62,"just":"Toilet plume studies indicate aerosol release."}
{"agent_id":"A2","t":0,"y":1,"p":0.58,"just":"Evidence appears mixed."}
{"agent_id":"A3","t":0,"y":0,"p":0.65,"just":"Particle counts increase after flushing."}
{"agent_id":"A4","t":0,"y":1,"p":0.55,"just":"Effect size may be limited."}
{"agent_id":"A5","t":0,"y":0,"p":0.60,"just":"Fluid dynamics support upward plume formation."}
{"agent_id":"A6","t":0,"y":0,"p":0.63,"just":"Bioaerosol reports align with open-lid flushing."}
{"agent_id":"A7","t":0,"y":1,"p":0.57,"just":"Prior evidence seems inconclusive."}
\end{PromptBlock}

\noindent
\textit{Across rounds $t=1$ and $t=2$, local neighbor interactions gradually shift
initially skeptical agents toward the majority stance.}

\paragraph{Consensus ($t=3$).}
\begin{PromptBlock}
{"agent_id":"A1","t":3,"y":0,"p":0.74,"just":"Neighborhood fully aligned on plume evidence."}
{"agent_id":"A2","t":3,"y":0,"p":0.73,"just":"Sustained agreement justifies True."}
{"agent_id":"A3","t":3,"y":0,"p":0.75,"just":"Evidence remains consistent across rounds."}
{"agent_id":"A4","t":3,"y":0,"p":0.70,"just":"Neighbor data resolves prior uncertainty."}
{"agent_id":"A5","t":3,"y":0,"p":0.74,"just":"Convergent aerosol observations."}
{"agent_id":"A6","t":3,"y":0,"p":0.75,"just":"Supportive studies maintain True."}
{"agent_id":"A7","t":3,"y":0,"p":0.73,"just":"Cumulative local evidence confirms the claim."}
\end{PromptBlock}

% -----------------------------
\subsection{Complete Graph Distributed Consensus (6 Neighbors, Failure Case)}

\paragraph{Claim.}
\emph{When spiders sense danger, they run toward people for protection.}

\paragraph{Ground Truth.}
False.

\paragraph{Topology and Outcome.}
A complete graph with seven agents (six neighbors each).
Despite initial disagreement, the system reaches unanimous consensus
($y=0$, True) at round $t=3$, resulting in a confident but incorrect judgment.

\paragraph{Initial Judgments ($t=0$).}
\begin{PromptBlock}
{"agent_id":"A1","t":0,"y":1,"p":0.63,"just":"Spiders generally avoid humans; approach seems unlikely."}
{"agent_id":"A2","t":0,"y":0,"p":0.58,"just":"Some anecdotal accounts suggest refuge near large objects."}
{"agent_id":"A3","t":0,"y":1,"p":0.61,"just":"Typical response is retreat from vibration sources."}
{"agent_id":"A4","t":0,"y":0,"p":0.55,"just":"Movement toward stationary masses could reduce exposure."}
{"agent_id":"A5","t":0,"y":1,"p":0.60,"just":"Ethology literature emphasizes avoidance behavior."}
{"agent_id":"A6","t":0,"y":0,"p":0.57,"just":"Shelter-seeking may incidentally align with human location."}
{"agent_id":"A7","t":0,"y":1,"p":0.59,"just":"Available cues suggest retreat, not approach."}
\end{PromptBlock}

\noindent
\textit{Through rounds $t=1$ and $t=2$, repeated exposure to a plausible but weak
“nearest-cover” interpretation increases its perceived credibility, despite
limited empirical support.}

\paragraph{Erroneous Consensus ($t=3$).}
\begin{PromptBlock}
{"agent_id":"A1","t":3,"y":0,"p":0.80,"just":"Nearest-cover account explains apparent approach behavior."}
{"agent_id":"A2","t":3,"y":0,"p":0.72,"just":"Convergent explanations support True."}
{"agent_id":"A3","t":3,"y":0,"p":0.71,"just":"Consistent shelter-seeking behavior validates interpretation."}
{"agent_id":"A4","t":3,"y":0,"p":0.74,"just":"Airflow and vibration gradients provide plausible refuge."}
{"agent_id":"A5","t":3,"y":0,"p":0.79,"just":"Nearest-cover heuristic yields apparent protection-seeking."}
{"agent_id":"A6","t":3,"y":0,"p":0.75,"just":"Network agreement justifies the conclusion."}
{"agent_id":"A7","t":3,"y":0,"p":0.73,"just":"Cumulative reports support approach-for-protection behavior."}
\end{PromptBlock}

\noindent
\textit{This case illustrates how strong conformity pressure in highly connected
networks can amplify a coherent yet incorrect narrative, leading to confident
misclassification.}

\section{Formal Definition of Metrics}
\label{sec:formula}

\subsection{Centralized Aggregation Metrics}

Let $y^{\star} \in \{0,1\}$ denote the ground-truth label for a given claim, and let $y_c$ denote the final judgment issued by the central node (hub or root).

\paragraph{Central Accuracy (CA).}
Central Accuracy measures the correctness of the collective decision produced by the central node:
\begin{equation}
\mathrm{CA} = \mathbb{I}[y_c = y^{\star}],
\end{equation}
where $\mathbb{I}[\cdot]$ is the indicator function. CA directly reflects the reliability of hub-mediated aggregation and isolates the effect of central competence from peripheral diversity.

\paragraph{Peripheral Accuracy (PA).}
Let $\mathcal{P}$ denote the set of peripheral (non-central) agents. Peripheral Accuracy is defined as
\begin{equation}
\mathrm{PA} = \frac{1}{|\mathcal{P}|} \sum_{i \in \mathcal{P}} \mathbb{I}[y_i = y^{\star}],
\end{equation}
capturing the average individual-level correctness independent of the aggregation outcome.

\paragraph{Center--Periphery Consistency (CPC).}
To quantify immediate conformity to the central decision, we define
\begin{equation}
\mathrm{CPC} = \frac{1}{|\mathcal{P}|} \sum_{i \in \mathcal{P}} \mathbb{I}[y_i = y_c].
\end{equation}
CPC measures the extent to which peripheral agents align with the hub’s judgment, irrespective of correctness, thereby isolating conformity strength from accuracy.

\subsection{Distributed Consensus Metrics}

In Distributed Consensus, agents iteratively update their judgments over rounds $t = 0, \dots, T_{\max}$. Let $y_i^{(t)}$ denote agent $i$’s judgment at round $t$.

\paragraph{Final Accuracy (FA).}
Let $\hat{y}$ denote the group decision at termination, defined as the unanimous label if consensus is reached, or the majority label at $T_{\max}$ otherwise. Final Accuracy is given by
\begin{equation}
\mathrm{FA} = \mathbb{I}[\hat{y} = y^{\star}],
\end{equation}
evaluating whether the emergent collective outcome matches the ground truth.

\paragraph{Time-to-Consensus (TTC).}
Time-to-Consensus is defined as
\begin{equation}
\small
\mathrm{TTC} = \min \left\{ t : y_1^{(t)} = y_2^{(t)} = \cdots = y_N^{(t)} \right\},
\end{equation}
and is undefined if unanimity is not achieved within $T_{\max}$. TTC reflects the efficiency of convergence induced by network connectivity and conformity strength.

\paragraph{Conformity Index (CI).}
At each round $t$, the Conformity Index is defined as

\begin{equation}
\mathrm{CI}^{(t)} =
\max\left(
\begin{aligned}
&\frac{1}{N} \sum_{i=1}^{N} \mathbb{I}\!\left[y_i^{(t)} = 1\right],\\
&\frac{1}{N} \sum_{i=1}^{N} \mathbb{I}\!\left[y_i^{(t)} = 0\right]
\end{aligned}
\right)
\end{equation}

For binary judgments, $\mathrm{CI}^{(t)} \in [0.5, 1]$, where $\mathrm{CI}^{(t)} = 1$ indicates unanimity and $\mathrm{CI}^{(t)} = 0.5$ corresponds to maximal disagreement.

\paragraph{Average Conformity Index (ACI).}
To summarize conformity dynamics over time, we report the Average CI:
\begin{equation}
\mathrm{ACI} = \frac{1}{T_{\max}} \sum_{t=1}^{T_{\max}} \mathrm{CI}^{(t)},
\end{equation}
which captures the overall tendency toward alignment across interaction rounds, rather than only the terminal state.

\end{document}